\documentclass[letter]{aa}  
\usepackage{graphicx}
\usepackage{txfonts}
\usepackage{hyperref}
\hypersetup{colorlinks=true, linkcolor=blue, citecolor=blue,filecolor=blue,urlcolor=blue}
\usepackage{xcolor}
\usepackage{array,multirow}

\makeatletter
\renewcommand*\aa@pageof{, page \thepage{} of \pageref*{LastPage}}
\makeatother

\bibpunct{(}{)}{;}{a}{}{,} 

\usepackage{tikz}
\definecolor{lime}{HTML}{A6CE39}
\definecolor{deeppurple}{RGB}{54,10,60}
\DeclareRobustCommand{\orcidicon}{
        \begin{tikzpicture}
        \draw[lime, fill=lime] (0,0) 
        circle [radius=0.16] 
        node[white] {{\fontfamily{qag}\selectfont \tiny ID}};
        \draw[white, fill=white] (-0.0625,0.095) 
        circle [radius=0.007];
        \end{tikzpicture}
        \hspace{-2mm}}

\foreach \x in {A, ..., Z}{\expandafter\xdef\csname orcid\x
\endcsname{\noexpand\href{https://orcid.org/\csname orcidauthor\x\endcsname}{\noexpand\orcidicon}}}

\newcommand{\vsini}{\mbox{$v\sin i$}}

\newcommand{\Teff}{\mbox{$T_{\rm eff}$}}
\newcommand{\logg}{\mbox{$\log g$}}

\newcommand{\logQ}{\mbox{$\log Q$}}
\newcommand{\logMdot}{\mbox{$\log \dot{M}$}}
\newcommand{\vinf}{$v_{\infty}$}
\newcommand{\vesc}{$v_{\rm esc}$}

\newcommand{\logL}{$\log (L/L_{\odot})$}

\let\oldAA\AA
\renewcommand*{\AA}{\,\oldAA\xspace}
\newcommand{\kms}{\,\mbox{km\,s$^{-1}$}\xspace}
\newcommand{\MSol}{\,\mbox{M$_\odot$}\xspace}

\newcommand{\ls}{\mbox{$\lesssim$}\,}
\newcommand{\gs}{\mbox{$\gtrsim$}\,}

\begin{document}

\title{The IACOB project}
\subtitle{XI. No increase in mass-loss rates over the bistability region}
\titlerunning{No increase in mass-loss rates over the bistability region}
\author{de~Burgos, A.\inst{1,2\orcidA{}}, 
Keszthelyi, Z.\inst{3\orcidC{}}, 
Simón-Díaz, S.\inst{1,2\orcidB{}}, 
Urbaneja, M.~A.\inst{4\orcidE{}}
}
\authorrunning{A. de~Burgos et al.}
\institute{
Universidad de La Laguna, Dpto. Astrof\'isica, E-38206 La Laguna, Tenerife, Spain
\and
Instituto de Astrof\'isica de Canarias, Avenida V\'ia L\'actea, E-38205 La Laguna, Tenerife, Spain
\and
Center for Computational Astrophysics, Division of Science, National Astronomical Observatory of Japan, 2-21-1, Osawa, Mitaka, Tokyo 181-8588, Japan
\and
Universit\"at Innsbruck, Institut f\"ur Astro- und Teilchenphysik, Technikerstr. 25/8, A-6020 Innsbruck, Austria
}
\date{Received 9 April 2024 / Accepted 17 May 2024}
\abstract 
{The properties of blue supergiants are key for constraining the end of the main sequence (MS) of massive stars. 
Whether the observed drop in the relative number of fast-rotating stars below $\approx$21\,kK is due to enhanced mass-loss rates at the location of the bistability jump, or the result of the end of the MS is still debated.
Here, we combine newly derived estimates of photospheric and wind parameters with \textit{Gaia} distances and wind terminal velocities from the literature to obtain upper limits on the mass-loss rates for a sample of 116 Galactic luminous blue supergiants.
The parameter space covered by the sample ranges between 35\,--\,15\,kK in \Teff\ and 4.8\,--\,5.8\,dex in \logL. 
Our results show no increase in the mass-loss rates over the bistability jump. Therefore, we argue that the drop in rotational velocities cannot be explained by enhanced mass loss.
Since a large jump in the mass-loss rates is commonly included in evolutionary models, we suggest an urgent revision of the  default prescriptions currently in use.
}

\maketitle


\section{Introduction}
\label{section.1_tmp}

Stellar winds play an important role in the evolution of massive stars with $M_{\rm ini}$\,\gs8\MSol \citep[e.g.,][for a review]{puls08}. Since winds are capable of driving a significant amount of mass into the interstellar medium, they can also alter
stellar lifetimes and, ultimately, the final fates of massive stars \citep[see][]{Maeder87, maeder09, smith14}.

The stellar wind models developed by \citet{pauldrach90} to investigate the massive star P Cygni revealed a bistable nature in the predicted wind properties of blue supergiants (BSGs). They identified a strong discontinuity in the mass-loss rates ($\dot{M}$) when the stellar parameters were only slightly modified. 
\citet{lamers95} showed the existence of a steep transition from high to low wind terminal velocities (\vinf) by a factor of 2 at $\approx$21\,kK. The decrease in \vinf\ was predicted to lead to an increase in the mass-loss rates also by a factor of 2, so that the wind momentum ($\dot{M}$\vinf) would remain constant over the bistability region.
This phenomenon is commonly referred to as the ``bistability jump."

Although the physical origin of the jump has been attributed to the recombination of Fe\,{\sc iv} into Fe\,{\sc iii} lines \citep[see][]{vink99}, the associated increase in the mass-loss rates remains unverified in statistically significant observational samples \citep[see][]{crowther06, markova08}, as noted in the recent review by \citet{vink22}.
Although theoretical calculations of \citet{vink01}, widely used in current evolutionary codes, predict an increase in the mass-loss rates by a factor of 5-7 over the bistability region at $\approx$25\,kK, more recent simulations carried out by \citet{krticka21} only reported an increase by a factor of 5 at $\approx$15\,kK. Furthermore, \citet{bjorklund21} did not predict such an increase at all.

Mass loss also leads to angular momentum loss during the evolution of massive stars \citep{langer98}. Therefore, enhanced mass loss, as predicted by \citet[][]{vink01}, can significantly spin down the star, leading to a ``bistability braking" \citep[][]{vink10}.
\citet{keszthelyi17} investigated the extent to which different scenarios of mass loss result in surface braking, suggesting that either additional mechanisms are required, or that the adopted initial rotation rates must be lower to explain the observed rotational velocities of BSGs.

In this letter, we present evidence of no enhanced mass loss over the bistability region (or at any temperature within 30\,--\,15\,kK) for Galactic stars in the mass range between 20 and 60\MSol. 
We compare our results with recent prescriptions of \citet{krticka21} and \citet{bjorklund23} to check which can better reproduce the observations. We also discuss the consequences for stellar evolution, in particular, in relation to the observed drop in the projected rotational velocity of BSGs and its connection to the end of the main sequence.


\section{Sample, data, and models}
\label{section.2}

\begin{figure}[!t]
  \centering
    \includegraphics[width=0.48\textwidth]{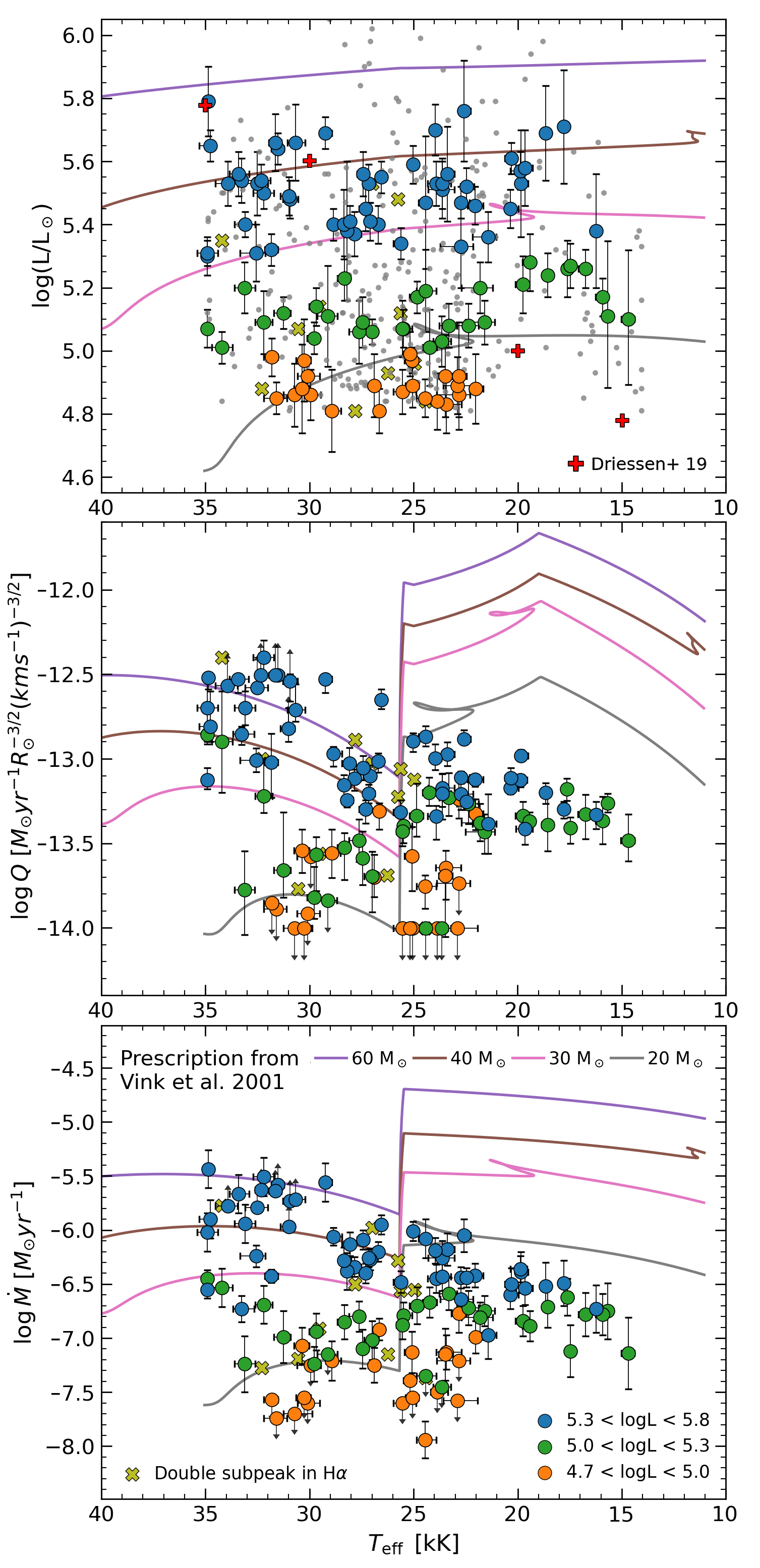}
    \caption{Properties of the 116 investigated Galactic stars described in Sect.~\ref{section.2}. 
    \emph{Top panel:} Location of the sample stars in the HR diagram. An additional 290 stars (without available UV spectroscopy) from \citet{deBurgos24a} are also shown for reference.
    \emph{Middle and bottom panels:} Wind-strength parameter (middle) and mass-loss rates (bottom) as a function of the effective temperature. Derived mass-loss rates were obtained using unclumped \textsc{FASTWIND} models. 
    In all of the panels, different colors separate stars in three luminosity ranges, as indicated in the legend.
    Upper and lower limits of $Q$ and $\dot{M}$ are indicated with arrows instead of error-bars.    
    Four evolutionary models with different initial masses and using the \citet{vink01} mass-loss prescription are included for reference.
    We note the predicted jump in $\dot{M}$ at $\approx$25\,kK.
    Stars showing double sub-peak emission in the H$\alpha$ line are indicated with lime-colored crosses and are discussed in the Appendix~\ref{apen.stellar_param}.
    The red plus symbols correspond to four models by \citet{driessen19} used to evaluate the clumping effects (see further explanation in Sect.~\ref{subsection.31_clumping}).}
    \label{fig.fig1}
\end{figure}

Our sample comprises 116 BSGs with wind terminal velocities available in the literature, spanning 35\,--\,15\,kK in \Teff\ and 4.8\,--\,5.8\,dex in \logL.
They were selected from the sample of 527 late-O to mid-B luminous blue stars analyzed by \citet{deBurgos24a} to cover $\approx$10\,kK on both sides of the bistability region following \citet{vink01}.
Their location in the Hertzsprung-Russell (HR) diagram is shown with filled circles in the top panel of Fig.~\ref{fig.fig1}, color-coded according to their luminosity.

Estimates of the effective temperature (\Teff), surface gravity (\logg), and the wind-strength parameter ($Q$) were obtained by means of quantitative spectroscopic analysis, employing models calculated with the \textsc{FASTWIND} code \citep{santolaya-rey97, puls05, rivero-gonzalez11, puls20}, as described in \citet{deBurgos24a}.
The wind strength parameter is defined as $Q = \dot{M} / (R_{\star}v_{\infty})^{1.5}$, where the mass-loss rate is in units of M$_\odot$~yr$^{-1}$, the wind terminal velocity is in units of \kms, and $R_{\star}$ is the stellar radius in units of $R_\odot$ \citep[see also][]{puls96, puls05}. For our analysis of optical spectra, the determination of $Q$ is mainly constrained by the H$\alpha$ profile. In addition, we note that we used the smooth wind option implemented in \textsc{FASTWIND}.

We then benefited from reliable estimates of \textit{Gaia} EDR3 distances \citep[with error-over-parallax values \ls0.15; see][]{bailer-jones21} and our own estimations of the total V-band extinction $(A_{\rm V})$ to derive the stellar radii and luminosities. Regarding \vinf, the adopted values were obtained from single-exposure UV spectra from \textit{IUE} (R\,=\,10\,000). About 93\% of the values are taken from the works of \citet{howarth89, howarth97}. Other values were adopted from \citet{prinja02, prinja10} (4\%) and \citet{lamers95} (3\%).
Using $Q$, $R_{\star}$, and \vinf, we derived $\dot{M}$ for the stars in the sample (see Sect.~\ref{section.3_massloss}). For further details, see Appendices~\ref{apen.stellar_param} and \ref{apen.uncertainties}.

We used the \textsc{mesa} tool \citep{paxton11, paxton13, paxton15, paxton18, paxton19} to compute one-dimensional (1D) stellar evolution models with initial masses from 20 to 60\,\MSol, from the zero-age main sequence (ZAMS) until 10\,kK. The models include rotation, diffusive angular momentum transport, overshooting, and mass loss. 
We compared three theoretical mass-loss prescriptions suitable for Galactic BSGs. First, we used \textsc{mesa}'s default routine, adopting the \citet{vink01} formula. Then, we implemented the mass-loss prescriptions from \citet{krticka21} and \citet{bjorklund23}. Further notes on the computed models are detailed in Appendix~\ref{apen.evo}.


\section{No increase in the mass-loss rates over the bistability region}
\label{section.3_massloss}

The middle and bottom panels of Fig.~\ref{fig.fig1} show our derived wind-strength $Q$-parameter values and mass-loss rates, respectively, against the effective temperature for our sample of stars.
Again, we remark that our methodology for deriving these two quantities did not account for wind clumping effects. However, while the impact of the presence of wind inhomogeneities is discussed in Sect.~\ref{subsection.31_clumping}, we anticipate that such effects do not change our main results.

The panels also include evolutionary model computations (see Sect.~\ref{section.2}) for four different initial masses, using the mass-loss prescription of \citet{vink01}. 
We should note that we did not anticipate self-consistent matches between the models and observations on these diagrams due to various issues\footnote{Firstly, the identification of a given evolutionary track that could correspond to a given star requires resolving the mass-discrepancy problem. Second, the observations do not include clumping corrections; hence, the values in $Q$ and $\dot{M}$ might shift downwards.}.
Rather, we consider that the majority of the stars in the sample follow a canonical evolutionary channel represented by these models, evolving from high to low effective temperatures. However, we cannot rule out the possibility that some of our objects may have followed a post-MS evolution or result from binary interaction. We investigated the helium enrichment of the O-stars and found that the majority are helium normal.
Given the flattening of the mass-luminosity relation for higher masses and the fact that for a given initial mass, stars evolve at an approximately constant luminosity in the range of 30\,--\,10\,kK, we decided to separate our sample into three luminosity bins (as indicated in the legend).

The prediction of \citet{vink01}, which is the most widely used prescription in stellar evolution modeling \citep[e.g.,][]{brott11, ekstrom12, paxton13}, results in a substantial increase in $\dot{M}$ by a factor of 10-20, when crossing the bistability region at $\approx$25\,kK\footnote{For our considered range of luminosities, the dependence of $\dot{M}$ with the Eddington factor as predicted in \citet{vink01} only shifts the position of the bistability by 1\,--\,2\,kK.} \citep[see also,][]{keszthelyi17}.
In contrast, our results for stars with \logL\ between 5.3 and 5.8\,dex (blue circles) show no evidence of a sudden increase in $\dot{M}$ (or $Q$) in this region. Instead, the observed stars within this luminosity range show a decrease in $\dot{M}$ by a factor of 10 from the hottest to the coolest objects.

Furthermore, stars with \logL\ between 5.0 and 5.3\,dex (green circles) display a ``quasi-flat" distribution in $\dot{M}$, again showing no evidence of the predicted increase in $\dot{M}$ (or $Q$) on the cool side of the bistability region.
The number of stars with \logL\ between 4.7 and 5.0\,dex (orange circles) is lower than those with higher luminosities. For half of this subgroup of stars, $Q$ (hence, also $\dot{M}$) is an upper limit. Nonetheless, we cannot identify a jump in mass-loss rates around 25\,kK.

We summarize the average values of $\dot{M}$ and $Q$ in Table~\ref{tab.avgQMdot} of Appendix~\ref{apen.meanval}. We conclude that in the luminosity bins considered, which approximately correspond to an evolutionary track with a given initial mass, there is no evidence of an increase in $\dot{M}$, as predicted by \citet{vink01}.


\subsection{The effect of clumping}
\label{subsection.31_clumping}

Due to the unstable nature of the wind-driving mechanism, the development of inhomogeneities (spatial and temporal) in the form of over- and underdense regions within the mass outflows of massive stars, has been postulated for several decades \citep[e.g.,][]{owocki88, sundqvist13}. On the observational side, several phenomena detected in different parts of the electromagnetic spectrum so far seem to support the presence of such inhomogeneities \citep[e.g.,][]{kaufer96, markova05, fullerton06, cohen2006, prinja2013}.

In the limiting case of the micro-clumping formalism, it is assumed that the optically thin over-dense clumps are surrounded by a void interclump medium. The density of the clumps is given by $\rho_\mathrm{cl}=f_\mathrm{cl}\,\rho$, with $f_\mathrm{cl}$ as the clumping factor and $\rho$ as the mean density. Using homogeneous wind models in the present work means that the mass-loss rates obtained from the analysis of H$\alpha$ represent a maximum value. This is the case for O-type stars \citep[see, e.g.,  ][]{sundqvist18}.

However, for late B-type supergiants (not covered in this work), H$\alpha$ should become (more) optically thick due to the partial recombination of hydrogen. As a consequence of the increased population of the lower level of the transition \citep[see][]{kudritzki00}, the H$\alpha$ line can potentially resemble the behavior of resonance lines in the UV spectrum of O-type stars \citep[becoming sensitive to porosity effects in velocity space; see][]{owocki08}.
Between these two limiting cases, H$\alpha$ should show a progressive change in character when crossing the B-type domain from early-to-late types \citep[see the discussion from][]{petrov14}. Therefore, we might naturally ask to what extent our results could be a consequence of considering homogeneous winds for the analyses.

We first consider the recent work by \citet{driessen19}, who employed wind instability simulations to investigate the formation of structures in the winds of typical O- and B-type supergiant stars. While these authors do not predict mass-loss rates in their simulations (and therefore we cannot compare our results in absolute terms), we can nonetheless consider the predicted growth rate of structures in the stellar wind.
In particular, \citet{driessen19} concluded that a significant difference in the properties of wind clumping should occur between the extremes of the region considered here, with O-stars developing significantly more clumping than the B-stars. Applying (somewhat naively) the clumping factors obtained by \citet{driessen19} in the formation region of H$\alpha$ to our sample, under the assumption that the clumps remain optically thin, would result in a reduction in our H$\alpha$-based mass-loss rates by a factor of 4 for the O-stars while they would remain basically unaffected for B-type supergiants. That is to say, the distribution of $\dot{M}$ and $Q$ values shown in Fig.~\ref{fig.fig1} would become flatter, but it would not present a "bistability jump-like" feature.

Utilizing the continuum emission from the far infrared to the radio regime, \citet{puls06} and \citet{rubio-diez22} studied the radial distribution of the clumping factors in the winds of O-type and B-type stars, respectively. Their results can be used to compare the behavior of the clumping factors up to 2R$_\star$, which is the relevant region for our H$\alpha$-derived mass-loss rates. The maximum clumping factors obtained for O-type stars are two to four times higher than those derived for B-type supergiants. Given that these results were obtained based on the assumption that the wind becomes homogeneous again in the radio region (the derived clumping factors should be re-scaled if this is not true), these values cannot be applied in absolute terms. However, taken in relative terms, they would suggest that there is not a sudden increase in $\dot{M}$ as a function of effective temperature within the parameter range studied here\footnote{We note here that, strictly, this would only remain true when the clumping properties in the radio-emitting region are the same for both types of object.}.

A refined treatment of wind inhomogeneities in model atmosphere codes (so-called macro-clumping) considers a wind composed of clumps (that can become optically thick) and an interclump medium that is not void \citep{oskinova07, sundqvist18}. The consideration of such macro-clumping for the analysis of samples of O- and B-type stars remains a novelty, with the works of \citet{hawcroft21} and \citet{brands22} pioneering the endeavor for O-type stars. These works show that a better agreement is obtained when resonance UV profiles are fitted under the macro-clumping formalism of \citet{sundqvist14}, while at the same time having little effect on H$\alpha$ (as expected). With a much more limited scope, \citet{bernini-peron23} illustrated the effect that the macro-clumping formalism of \citet{oskinova07} has on the Galactic B3\,Ia star HD\,53138. 

We performed a set of simulations with \textsc{FASTWIND} to assess how the incorporation of macro-clumping, using the \cite{sundqvist18} formalism, would affect our results. We stress here that under this formalism, the clumps are not assumed to be optically thin or thick; rather, the optical depth is calculated using the Sobolev approximation \citep{sobolev60}. The effects of velocity-porosity are incorporated as well. With this effective-opacity treatment, the rate equations and the ionization and excitation equilibria can change accordingly. Further details of these simulations are provided in Appendix~\ref{apen.macroclump}. The main conclusion of this exercise is that for the BSG domain studied in the present work when considering a sensible set of parameters describing the clumps, macro-clumping does not modify the mass-loss rates determined from the H$\alpha$ line profiles. Therefore, the trends identified in $\dot{M}$ and $Q$ are not altered. This result is qualitatively in line with the outcome of the 3D Monte Carlo simulations performed by \citet{surlan12}.


\begin{figure}[!t]
    \centering
    \includegraphics[width=0.475\textwidth]{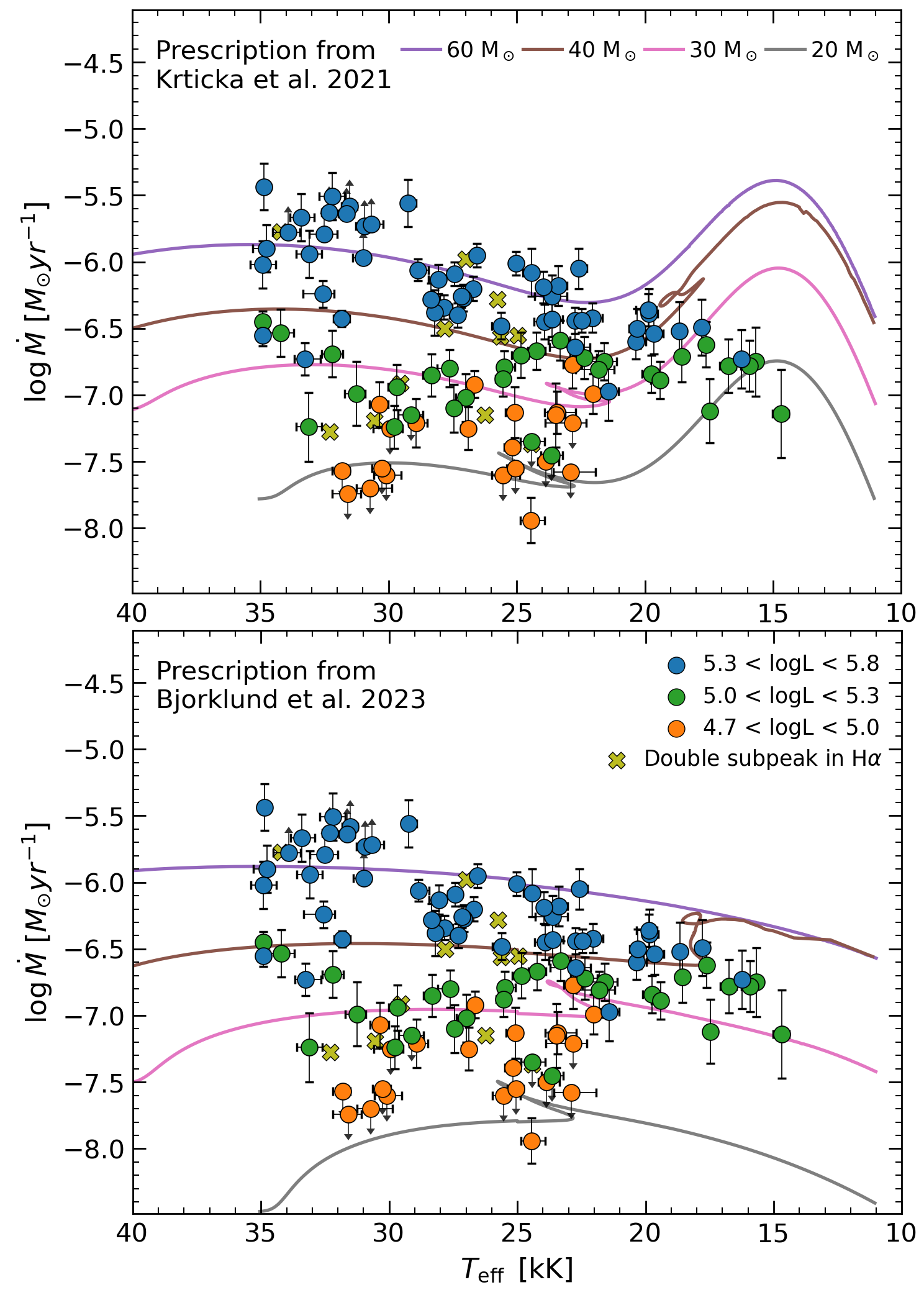}
    \caption{Mass-loss rates against effective temperature for two additional mass-loss prescriptions. The top panel follows \citet{krticka21} and the bottom panel follows \citet{bjorklund23}. Each panel includes the same set of colors and symbols as in Fig.~\ref{fig.fig1}.}
    \label{fig.fig2}
\end{figure}

\section{Implications for massive star evolution}
\label{section.4_evolution}


\subsection{Alternative prescriptions of theoretical wind models}
\label{subsection.41_prescriptions}

While theoretical predictions of mass-loss rates from \citet{vink01} have been used for several years in stellar evolutionary model computations, our results suggest the need for alternative prescriptions \citep[see also][]{crowther06, markova08, keszthelyi17, rubio-diez22}. In this regard, we illustrate with Fig.~\ref{fig.fig2} the mass-loss rates against the effective temperature for evolutionary models adopting the mass-loss rates of \citet{krticka21} and \citet{bjorklund23} (upper and lower panels, respectively). Both show a similar behavior from 40 to 22\,kK by gradually decreasing mass-loss rate with \Teff. These prescriptions are consistent with the trends displayed in our data, showing no increase within the 30\,--\,20\,kK range. The rates by \citet{krticka21} include a gradual rise below 22\,kK, with a peak at $\approx$15\,kK, whereas the rates by \citet{bjorklund23} continue to decrease below that temperature. The number of stars near 15\,kK is still small to draw any firm conclusion.


\subsection{A connection between the extension of the main sequence and the spin-rate properties of BSGs}
\label{subsection.42_ms-end}

\begin{figure}[!t]
    \centering
    \includegraphics[width=0.485\textwidth]{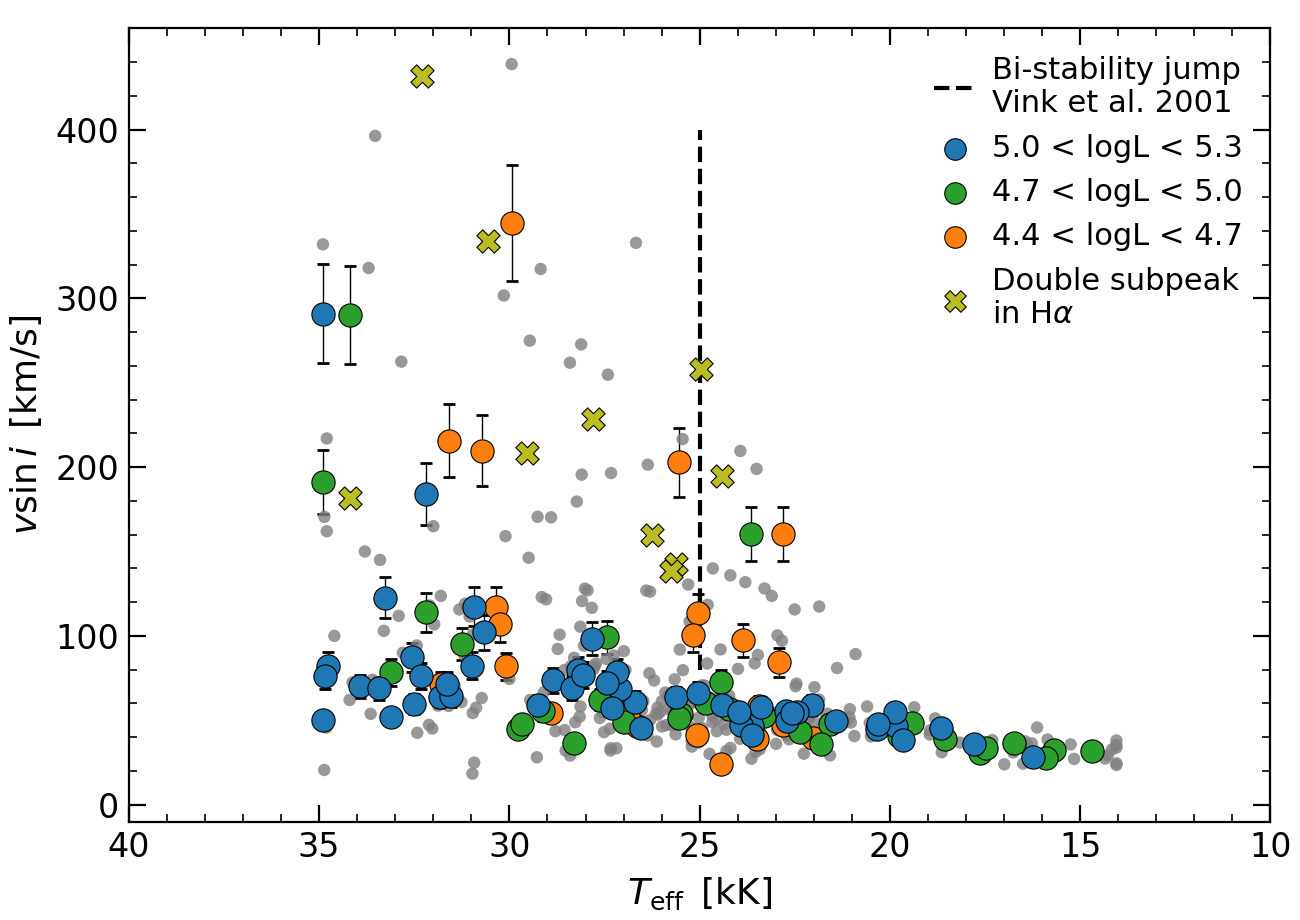}
    \caption{Estimates of the projected rotational velocity against the derived effective temperatures. Symbols and colors are the same as in the bottom panel of Fig.~\ref{fig.fig1}. The dashed line shows the effective temperature where the increase in $\dot{M}$ takes place, as implemented in \textsc{mesa} (see Appendix \ref{apen.evo}).}
    \label{fig.vsini}
\end{figure}

Whether the majority of BSGs are main sequence (MS) or post-MS objects remains debated \citep[e.g.,][]{vink10, castro14}. As a result of the rapid evolution toward cooler temperatures, we would expect the relative number of post-MS objects to be much lower compared to their MS counterparts. Additionally, post-MS objects are expected to undergo a significant decrease in their surface rotational velocity due to the significant increase in their size.

The lack of fast-rotating stars with projected rotational velocities (\vsini) greater than $\approx$100\kms below a given temperature has been known for more than three decades \citep[e.g.,][]{howarth97, ryans02, vink10, simon-diaz14a, mcevoy15, deburgos23a}. We demonstrate this with our sample stars in Fig.~\ref{fig.vsini}, finding a drop at $\approx$21\,kK.
As indicated by \citet{vink10}, this feature could be related to the termination of the MS \citep[see also][]{brott11}. In fact, the observed drop appears to overlap with the drop in the relative number of stars, as highlighted by \citet[][]{deBurgos24a}. However, as an alternative explanation, \citet{vink10} suggested that increased mass-loss rates at the bistability jump can trigger ``bistability braking," leading to the spin-down of stars.
In this work, we find no evidence for increased mass-loss rates in the bistability region. Consequently, we argue that the lack of fast-rotating stars is more likely connected with the terminal-age main sequence (TAMS). 
We also point out that for low-\vsini\ objects, which comprise the majority of BSGs \citep{deBurgos24a}, there is no evidence of a drop in \vsini\ around the bi-stability region, thus favoring the above possibility once again.
Nevertheless, the study of the location of the TAMS requires further considerations. These include, among other things, the analysis of much larger samples, an evaluation of potential observational biases, the inclusion of hotter O-type stars, and the consideration of additional evolutionary channels populating the HR diagram, such as binary products \citep{demink14} or post-red supergiants that have undergone a blue-loop evolution \citep{martinet21}.


\section{Final remarks}
\label{section.5_conclusions}

The results of this work add strong constraints to the mass-loss rates of BSGs, providing a reassessment of its behavior in the 35\,--\,15\,kK effective temperature range. We chose this range to generously cover $\approx$25\,kK, where \citet{vink01} predicted a significant increase in mass-loss rates. To date, this increase is widely implemented in most evolutionary codes \citep[e.g.,][]{brott11, ekstrom12, paxton13}, but alternatives have been tested \citep[e.g.,][]{keszthelyi17}.

Although a better characterization of the clumping properties in the whole BSG domain can lead to a more robust description of the dependence of mass-loss rates with effective temperature and luminosity, our study provides strong observational evidence of the lack of an increase in mass-loss rates over the bistability region. As a consequence, from the proposed explanations for the absence of fast-rotating stars below a certain effective temperature, the most plausible one would be the termination of the main sequence.\\

As potential areas for improvement in the future, the simultaneous fitting of optical and UV spectra for our complete sample will provide additional clues on mass-loss rates. We also emphasize the importance of increasing the number of stars with available wind terminal velocities from UV spectra, as well as exploring mass-loss rates for stars with effective temperatures below 15\,kK. In particular, the latter will help to provide additional constraints to the newer prescriptions of \citet{krticka21} and \citet{bjorklund21}.


\begin{acknowledgements}

We thank the anonymous reviewer for a constructive and positive report that has led to the improvement of the manuscript.
We are also deeply indebted to Joachim Puls for providing valuable comments on this manuscript and for his continuous support.
AdB and SS-D acknowledge support from the Spanish Ministry of Science and Innovation (MICINN) through the Spanish State Research Agency through grants PID2021-122397NB-C21, and the Severo Ochoa Programme 2020-2023 (CEX2019-000920-S).
ZK acknowledges support from the Overseas Visit Program for Young Researchers from the National Astronomical Observatory of Japan and the Early-Career Visitor Program from the Instituto de Astrof\'isica de Canarias. 
Numerical computations were carried out on the general-purpose PC cluster at the Center for Computational Astrophysics, National Astronomical Observatory of Japan.

\end{acknowledgements}

\bibliographystyle{aa} 
\bibliography{biblio} 

\begin{thebibliography}{83}
\expandafter\ifx\csname natexlab\endcsname\relax\def\natexlab#1{#1}\fi

\bibitem[{{Asplund} {et~al.}(2009){Asplund}, {Grevesse}, {Sauval}, \&
  {Scott}}]{asplund09}
{Asplund}, M., {Grevesse}, N., {Sauval}, A.~J., \& {Scott}, P. 2009, \araa, 47,
  481

\bibitem[{{Bailer-Jones} {et~al.}(2021){Bailer-Jones}, {Rybizki}, {Fouesneau},
  {Demleitner}, \& {Andrae}}]{bailer-jones21}
{Bailer-Jones}, C.~A.~L., {Rybizki}, J., {Fouesneau}, M., {Demleitner}, M., \&
  {Andrae}, R. 2021, \aj, 161, 147

\bibitem[{{Bernini-Peron} {et~al.}(2023){Bernini-Peron}, {Marcolino}, {Sander},
  {Bouret}, {Ramachandran}, {Saling}, {Schneider}, {Oskinova}, \&
  {Najarro}}]{bernini-peron23}
{Bernini-Peron}, M., {Marcolino}, W.~L.~F., {Sander}, A.~A.~C., {et~al.} 2023,
  \aap, 677, A50

\bibitem[{{Bj{\"o}rklund} {et~al.}(2021){Bj{\"o}rklund}, {Sundqvist}, {Puls},
  \& {Najarro}}]{bjorklund21}
{Bj{\"o}rklund}, R., {Sundqvist}, J.~O., {Puls}, J., \& {Najarro}, F. 2021,
  \aap, 648, A36

\bibitem[{{Bj{\"o}rklund} {et~al.}(2023){Bj{\"o}rklund}, {Sundqvist}, {Singh},
  {Puls}, \& {Najarro}}]{bjorklund23}
{Bj{\"o}rklund}, R., {Sundqvist}, J.~O., {Singh}, S.~M., {Puls}, J., \&
  {Najarro}, F. 2023, \aap, 676, A109

\bibitem[{{B{\"o}hm-Vitense}(1958)}]{bohm58}
{B{\"o}hm-Vitense}, E. 1958, \zap, 46, 108

\bibitem[{{Brands} {et~al.}(2022){Brands}, {de Koter}, {Bestenlehner},
  {Crowther}, {Sundqvist}, {Puls}, {Caballero-Nieves}, {Abdul-Masih},
  {Driessen}, {Garc{\'\i}a}, {Geen}, {Gr{\"a}fener}, {Hawcroft}, {Kaper},
  {Keszthelyi}, {Langer}, {Sana}, {Schneider}, {Shenar}, \& {Vink}}]{brands22}
{Brands}, S.~A., {de Koter}, A., {Bestenlehner}, J.~M., {et~al.} 2022, \aap,
  663, A36

\bibitem[{{Brott} {et~al.}(2011){Brott}, {de Mink}, {Cantiello}, {Langer}, {de
  Koter}, {Evans}, {Hunter}, {Trundle}, \& {Vink}}]{brott11}
{Brott}, I., {de Mink}, S.~E., {Cantiello}, M., {et~al.} 2011, \aap, 530, A115

\bibitem[{{Castro} {et~al.}(2014){Castro}, {Fossati}, {Langer},
  {Sim{\'o}n-D{\'\i}az}, {Schneider}, \& {Izzard}}]{castro14}
{Castro}, N., {Fossati}, L., {Langer}, N., {et~al.} 2014, \aap, 570, L13

\bibitem[{{Cohen} {et~al.}(2006){Cohen}, {Leutenegger}, {Grizzard}, {Reed},
  {Kramer}, \& {Owocki}}]{cohen2006}
{Cohen}, D.~H., {Leutenegger}, M.~A., {Grizzard}, K.~T., {et~al.} 2006, \mnras,
  368, 1905

\bibitem[{{Crowther} {et~al.}(2006){Crowther}, {Lennon}, \&
  {Walborn}}]{crowther06}
{Crowther}, P.~A., {Lennon}, D.~J., \& {Walborn}, N.~R. 2006, \aap, 446, 279

\bibitem[{{de Almeida} {et~al.}(2019){de Almeida}, {Marcolino}, {Bouret}, \&
  {Pereira}}]{deAlmeida19}
{de Almeida}, E.~S.~G., {Marcolino}, W.~L.~F., {Bouret}, J.~C., \& {Pereira},
  C.~B. 2019, \aap, 628, A36

\bibitem[{{de Burgos} {et~al.}(2023{\natexlab{a}}){de Burgos},
  {Sim{\'o}n-D{\'\i}az}, {Urbaneja}, \& {Negueruela}}]{deburgos23a}
{de Burgos}, A., {Sim{\'o}n-D{\'\i}az}, S., {Urbaneja}, M.~A., \& {Negueruela},
  I. 2023{\natexlab{a}}, \aap, 674, A212

\bibitem[{{de Burgos} {et~al.}(2023{\natexlab{b}}){de Burgos},
  {Sim{\'o}n-D{\'\i}az}, {Urbaneja}, \& {Puls}}]{deBurgos24a}
{de Burgos}, A., {Sim{\'o}n-D{\'\i}az}, S., {Urbaneja}, M.~A., \& {Puls}, J.
  2023{\natexlab{b}}, arXiv e-prints, arXiv:2312.00241

\bibitem[{{de Mink} {et~al.}(2014){de Mink}, {Sana}, {Langer}, {Izzard}, \&
  {Schneider}}]{demink14}
{de Mink}, S.~E., {Sana}, H., {Langer}, N., {Izzard}, R.~G., \& {Schneider},
  F.~R.~N. 2014, \apj, 782, 7

\bibitem[{{Driessen} {et~al.}(2019){Driessen}, {Sundqvist}, \&
  {Kee}}]{driessen19}
{Driessen}, F.~A., {Sundqvist}, J.~O., \& {Kee}, N.~D. 2019, \aap, 631, A172

\bibitem[{{Ekstr{\"o}m} {et~al.}(2012){Ekstr{\"o}m}, {Georgy}, {Eggenberger},
  {Meynet}, {Mowlavi}, {Wyttenbach}, {Granada}, {Decressin}, {Hirschi},
  {Frischknecht}, {Charbonnel}, \& {Maeder}}]{ekstrom12}
{Ekstr{\"o}m}, S., {Georgy}, C., {Eggenberger}, P., {et~al.} 2012, \aap, 537,
  A146

\bibitem[{{Fullerton} {et~al.}(2006){Fullerton}, {Massa}, \&
  {Prinja}}]{fullerton06}
{Fullerton}, A.~W., {Massa}, D.~L., \& {Prinja}, R.~K. 2006, \apj, 637, 1025

\bibitem[{{Gaia Collaboration} {et~al.}(2021){Gaia Collaboration}, {Brown},
  {Vallenari}, {Prusti}, {de Bruijne}, {Babusiaux}, {Biermann}, {Creevey},
  {Evans}, {Eyer}, {Hutton}, {Jansen}, {Jordi}, {Klioner}, {Lammers},
  {Lindegren}, {Luri}, {Mignard}, {Panem}, {Pourbaix}, {Randich}, {Sartoretti},
  {Soubiran}, {Walton}, {Arenou}, {Bailer-Jones}, {Bastian}, {Cropper},
  {Drimmel}, {Katz}, {Lattanzi}, {van Leeuwen}, {Bakker}, {Cacciari},
  {Casta{\~n}eda}, {De Angeli}, {Ducourant}, {Fabricius}, {Fouesneau},
  {Fr{\'e}mat}, {Guerra}, {Guerrier}, {Guiraud}, {Jean-Antoine Piccolo},
  {Masana}, {Messineo}, {Mowlavi}, {Nicolas}, {Nienartowicz}, {Pailler},
  {Panuzzo}, {Riclet}, {Roux}, {Seabroke}, {Sordo}, {Tanga}, {Th{\'e}venin},
  {Gracia-Abril}, {Portell}, {Teyssier}, {Altmann}, {Andrae}, {Bellas-Velidis},
  {Benson}, {Berthier}, {Blomme}, {Brugaletta}, {Burgess}, {Busso}, {Carry},
  {Cellino}, {Cheek}, {Clementini}, {Damerdji}, {Davidson}, {Delchambre},
  {Dell'Oro}, {Fern{\'a}ndez-Hern{\'a}ndez}, {Galluccio}, {Garc{\'\i}a-Lario},
  {Garcia-Reinaldos}, {Gonz{\'a}lez-N{\'u}{\~n}ez}, {Gosset}, {Haigron},
  {Halbwachs}, {Hambly}, {Harrison}, {Hatzidimitriou}, {Heiter},
  {Hern{\'a}ndez}, {Hestroffer}, {Hodgkin}, {Holl}, {Jan{\ss}en}, {Jevardat de
  Fombelle}, {Jordan}, {Krone-Martins}, {Lanzafame}, {L{\"o}ffler}, {Lorca},
  {Manteiga}, {Marchal}, {Marrese}, {Moitinho}, {Mora}, {Muinonen}, {Osborne},
  {Pancino}, {Pauwels}, {Petit}, {Recio-Blanco}, {Richards}, {Riello},
  {Rimoldini}, {Robin}, {Roegiers}, {Rybizki}, {Sarro}, {Siopis}, {Smith},
  {Sozzetti}, {Ulla}, {Utrilla}, {van Leeuwen}, {van Reeven}, {Abbas}, {Abreu
  Aramburu}, {Accart}, {Aerts}, {Aguado}, {Ajaj}, {Altavilla}, {{\'A}lvarez},
  {{\'A}lvarez Cid-Fuentes}, {Alves}, {Anderson}, {Anglada Varela}, {Antoja},
  {Audard}, {Baines}, {Baker}, {Balaguer-N{\'u}{\~n}ez}, {Balbinot}, {Balog},
  {Barache}, {Barbato}, {Barros}, {Barstow}, {Bartolom{\'e}}, {Bassilana},
  {Bauchet}, {Baudesson-Stella}, {Becciani}, {Bellazzini}, {Bernet}, {Bertone},
  {Bianchi}, {Blanco-Cuaresma}, {Boch}, {Bombrun}, {Bossini}, {Bouquillon},
  {Bragaglia}, {Bramante}, {Breedt}, {Bressan}, {Brouillet}, {Bucciarelli},
  {Burlacu}, {Busonero}, {Butkevich}, {Buzzi}, {Caffau}, {Cancelliere},
  {C{\'a}novas}, {Cantat-Gaudin}, {Carballo}, {Carlucci}, {Carnerero},
  {Carrasco}, {Casamiquela}, {Castellani}, {Castro-Ginard}, {Castro Sampol},
  {Chaoul}, {Charlot}, {Chemin}, {Chiavassa}, {Cioni}, {Comoretto}, {Cooper},
  {Cornez}, {Cowell}, {Crifo}, {Crosta}, {Crowley}, {Dafonte}, {Dapergolas},
  {David}, {David}, {de Laverny}, {De Luise}, {De March}, {De Ridder}, {de
  Souza}, {de Teodoro}, {de Torres}, {del Peloso}, {del Pozo}, {Delbo},
  {Delgado}, {Delgado}, {Delisle}, {Di Matteo}, {Diakite}, {Diener},
  {Distefano}, {Dolding}, {Eappachen}, {Edvardsson}, {Enke}, {Esquej}, {Fabre},
  {Fabrizio}, {Faigler}, {Fedorets}, {Fernique}, {Fienga}, {Figueras},
  {Fouron}, {Fragkoudi}, {Fraile}, {Franke}, {Gai}, {Garabato},
  {Garcia-Gutierrez}, {Garc{\'\i}a-Torres}, {Garofalo}, {Gavras}, {Gerlach},
  {Geyer}, {Giacobbe}, {Gilmore}, {Girona}, {Giuffrida}, {Gomel}, {Gomez},
  {Gonzalez-Santamaria}, {Gonz{\'a}lez-Vidal}, {Granvik},
  {Guti{\'e}rrez-S{\'a}nchez}, {Guy}, {Hauser}, {Haywood}, {Helmi}, {Hidalgo},
  {Hilger}, {H{\l}adczuk}, {Hobbs}, {Holland}, {Huckle}, {Jasniewicz},
  {Jonker}, {Juaristi Campillo}, {Julbe}, {Karbevska}, {Kervella}, {Khanna},
  {Kochoska}, {Kontizas}, {Kordopatis}, {Korn}, {Kostrzewa-Rutkowska},
  {Kruszy{\'n}ska}, {Lambert}, {Lanza}, {Lasne}, {Le Campion}, {Le Fustec},
  {Lebreton}, {Lebzelter}, {Leccia}, {Leclerc}, {Lecoeur-Taibi}, {Liao},
  {Licata}, {Lindstr{\o}m}, {Lister}, {Livanou}, {Lobel}, {Madrero Pardo},
  {Managau}, {Mann}, {Marchant}, {Marconi}, {Marcos Santos}, {Marinoni},
  {Marocco}, {Marshall}, {Martin Polo}, {Mart{\'\i}n-Fleitas}, {Masip},
  {Massari}, {Mastrobuono-Battisti}, {Mazeh}, {McMillan}, {Messina},
  {Michalik}, {Millar}, {Mints}, {Molina}, {Molinaro}, {Moln{\'a}r},
  {Montegriffo}, {Mor}, {Morbidelli}, {Morel}, {Morris}, {Mulone}, {Munoz},
  {Muraveva}, {Murphy}, {Musella}, {Noval}, {Ord{\'e}novic}, {Orr{\`u}},
  {Osinde}, {Pagani}, {Pagano}, {Palaversa}, {Palicio}, {Panahi}, {Pawlak},
  {Pe{\~n}alosa Esteller}, {Penttil{\"a}}, {Piersimoni}, {Pineau}, {Plachy},
  {Plum}, {Poggio}, {Poretti}, {Poujoulet}, {Pr{\v{s}}a}, {Pulone}, {Racero},
  {Ragaini}, {Rainer}, {Raiteri}, {Rambaux}, {Ramos}, {Ramos-Lerate}, {Re
  Fiorentin}, {Regibo}, {Reyl{\'e}}, {Ripepi}, {Riva}, {Rixon}, {Robichon},
  {Robin}, {Roelens}, {Rohrbasser}, {Romero-G{\'o}mez}, {Rowell}, {Royer},
  {Rybicki}, {Sadowski}, {Sagrist{\`a} Sell{\'e}s}, {Sahlmann}, {Salgado},
  {Salguero}, {Samaras}, {Sanchez Gimenez}, {Sanna}, {Santove{\~n}a},
  {Sarasso}, {Schultheis}, {Sciacca}, {Segol}, {Segovia}, {S{\'e}gransan},
  {Semeux}, {Shahaf}, {Siddiqui}, {Siebert}, {Siltala}, {Slezak}, {Smart},
  {Solano}, {Solitro}, {Souami}, {Souchay}, {Spagna}, {Spoto}, {Steele},
  {Steidelm{\"u}ller}, {Stephenson}, {S{\"u}veges}, {Szabados}, {Szegedi-Elek},
  {Taris}, {Tauran}, {Taylor}, {Teixeira}, {Thuillot}, {Tonello}, {Torra},
  {Torra}, {Turon}, {Unger}, {Vaillant}, {van Dillen}, {Vanel}, {Vecchiato},
  {Viala}, {Vicente}, {Voutsinas}, {Weiler}, {Wevers}, {Wyrzykowski}, {Yoldas},
  {Yvard}, {Zhao}, {Zorec}, {Zucker}, {Zurbach}, \&
  {Zwitter}}]{gaiacollaboration21}
{Gaia Collaboration}, {Brown}, A.~G.~A., {Vallenari}, A., {et~al.} 2021, \aap,
  649, A1

\bibitem[{{Hawcroft} {et~al.}(2021){Hawcroft}, {Sana}, {Mahy}, {Sundqvist},
  {Abdul-Masih}, {Bouret}, {Brands}, {de Koter}, {Driessen}, \&
  {Puls}}]{hawcroft21}
{Hawcroft}, C., {Sana}, H., {Mahy}, L., {et~al.} 2021, \aap, 655, A67

\bibitem[{{Herrero} {et~al.}(1992){Herrero}, {Kudritzki}, {Vilchez}, {Kunze},
  {Butler}, \& {Haser}}]{herrero92}
{Herrero}, A., {Kudritzki}, R.~P., {Vilchez}, J.~M., {et~al.} 1992, \aap, 261,
  209

\bibitem[{{Higgins} {et~al.}(2021){Higgins}, {Sander}, {Vink}, \&
  {Hirschi}}]{higgins21}
{Higgins}, E.~R., {Sander}, A.~A.~C., {Vink}, J.~S., \& {Hirschi}, R. 2021,
  \mnras, 505, 4874

\bibitem[{{Howarth} \& {Prinja}(1989)}]{howarth89}
{Howarth}, I.~D. \& {Prinja}, R.~K. 1989, \apjs, 69, 527

\bibitem[{{Howarth} {et~al.}(1997){Howarth}, {Siebert}, {Hussain}, \&
  {Prinja}}]{howarth97}
{Howarth}, I.~D., {Siebert}, K.~W., {Hussain}, G. A.~J., \& {Prinja}, R.~K.
  1997, \mnras, 284, 265

\bibitem[{{Kaufer} {et~al.}(1996){Kaufer}, {Stahl}, {Wolf}, {Gaeng},
  {Gummersbach}, {Kovacs}, {Mandel}, \& {Szeifert}}]{kaufer96}
{Kaufer}, A., {Stahl}, O., {Wolf}, B., {et~al.} 1996, \aap, 305, 887

\bibitem[{{Keszthelyi} {et~al.}(2022){Keszthelyi}, {de Koter}, {G{\"o}tberg},
  {Meynet}, {Brands}, {Petit}, {Carrington}, {David-Uraz}, {Geen}, {Georgy},
  {Hirschi}, {Puls}, {Ramalatswa}, {Shultz}, \& {ud-Doula}}]{keszthelyi22}
{Keszthelyi}, Z., {de Koter}, A., {G{\"o}tberg}, Y., {et~al.} 2022, \mnras,
  517, 2028

\bibitem[{{Keszthelyi} {et~al.}(2017){Keszthelyi}, {Puls}, \&
  {Wade}}]{keszthelyi17}
{Keszthelyi}, Z., {Puls}, J., \& {Wade}, G.~A. 2017, \aap, 598, A4

\bibitem[{{Krti{\v{c}}ka} {et~al.}(2021){Krti{\v{c}}ka}, {Kub{\'a}t}, \&
  {Krti{\v{c}}kov{\'a}}}]{krticka21}
{Krti{\v{c}}ka}, J., {Kub{\'a}t}, J., \& {Krti{\v{c}}kov{\'a}}, I. 2021, \aap,
  647, A28

\bibitem[{{Kudritzki} \& {Puls}(2000)}]{kudritzki00}
{Kudritzki}, R.-P. \& {Puls}, J. 2000, \araa, 38, 613

\bibitem[{{Lamers} {et~al.}(1995){Lamers}, {Snow}, \& {Lindholm}}]{lamers95}
{Lamers}, H. J.~G.~L.~M., {Snow}, T.~P., \& {Lindholm}, D.~M. 1995, \apj, 455,
  269

\bibitem[{{Langer}(1998)}]{langer98}
{Langer}, N. 1998, \aap, 329, 551

\bibitem[{{Lodders}(2003)}]{lodders03}
{Lodders}, K. 2003, \apj, 591, 1220

\bibitem[{{Maeder}(2009)}]{maeder09}
{Maeder}, A. 2009, {Physics, Formation and Evolution of Rotating Stars}

\bibitem[{{Maeder} \& {Meynet}(1987)}]{Maeder87}
{Maeder}, A. \& {Meynet}, G. 1987, \aap, 182, 243

\bibitem[{{Maeder} \& {Meynet}(2000)}]{maeder00}
{Maeder}, A. \& {Meynet}, G. 2000, \araa, 38, 143

\bibitem[{{Marcolino} {et~al.}(2009){Marcolino}, {Bouret}, {Martins},
  {Hillier}, {Lanz}, \& {Escolano}}]{marcolino09}
{Marcolino}, W.~L.~F., {Bouret}, J.~C., {Martins}, F., {et~al.} 2009, \aap,
  498, 837

\bibitem[{{Markova} \& {Puls}(2008)}]{markova08}
{Markova}, N. \& {Puls}, J. 2008, \aap, 478, 823

\bibitem[{{Markova} {et~al.}(2005){Markova}, {Puls}, {Scuderi}, \&
  {Markov}}]{markova05}
{Markova}, N., {Puls}, J., {Scuderi}, S., \& {Markov}, H. 2005, \aap, 440, 1133

\bibitem[{{Martinet} {et~al.}(2021){Martinet}, {Meynet}, {Ekstr{\"o}m},
  {Sim{\'o}n-D{\'\i}az}, {Holgado}, {Castro}, {Georgy}, {Eggenberger},
  {Buldgen}, {Salmon}, {Hirschi}, {Groh}, {Farrell}, \& {Murphy}}]{martinet21}
{Martinet}, S., {Meynet}, G., {Ekstr{\"o}m}, S., {et~al.} 2021, \aap, 648, A126

\bibitem[{{Martins} {et~al.}(2005){Martins}, {Schaerer}, {Hillier},
  {Meynadier}, {Heydari-Malayeri}, \& {Walborn}}]{martins05b}
{Martins}, F., {Schaerer}, D., {Hillier}, D.~J., {et~al.} 2005, \aap, 441, 735

\bibitem[{{McEvoy} {et~al.}(2015){McEvoy}, {Dufton}, {Evans}, {Kalari},
  {Markova}, {Sim{\'o}n-D{\'\i}az}, {Vink}, {Walborn}, {Crowther}, {de Koter},
  {de Mink}, {Dunstall}, {H{\'e}nault-Brunet}, {Herrero}, {Langer}, {Lennon},
  {Ma{\'\i}z Apell{\'a}niz}, {Najarro}, {Puls}, {Sana}, {Schneider}, \&
  {Taylor}}]{mcevoy15}
{McEvoy}, C.~M., {Dufton}, P.~L., {Evans}, C.~J., {et~al.} 2015, \aap, 575, A70

\bibitem[{{Mermilliod}(2006)}]{mermilliod06}
{Mermilliod}, J.~C. 2006, {VizieR Online Data Catalog: Homogeneous Means in the
  UBV System (Mermilliod 1991)}, VizieR On-line Data Catalog: II/168.
  Originally published in: Institut d'Astronomie, Universite de Lausanne (1991)

\bibitem[{{Nieva} \& {Przybilla}(2012)}]{nieva12}
{Nieva}, M.~F. \& {Przybilla}, N. 2012, \aap, 539, A143

\bibitem[{{Oskinova} {et~al.}(2007){Oskinova}, {Hamann}, \&
  {Feldmeier}}]{oskinova07}
{Oskinova}, L.~M., {Hamann}, W.~R., \& {Feldmeier}, A. 2007, \aap, 476, 1331

\bibitem[{{Owocki}(2008)}]{owocki08}
{Owocki}, S.~P. 2008, in Clumping in Hot-Star Winds, ed. W.-R. {Hamann},
  A.~{Feldmeier}, \& L.~M. {Oskinova}, 121

\bibitem[{{Owocki} {et~al.}(1988){Owocki}, {Castor}, \& {Rybicki}}]{owocki88}
{Owocki}, S.~P., {Castor}, J.~I., \& {Rybicki}, G.~B. 1988, \apj, 335, 914

\bibitem[{{Pauldrach} \& {Puls}(1990)}]{pauldrach90}
{Pauldrach}, A.~W.~A. \& {Puls}, J. 1990, \aap, 237, 409

\bibitem[{{Paxton} {et~al.}(2011){Paxton}, {Bildsten}, {Dotter}, {Herwig},
  {Lesaffre}, \& {Timmes}}]{paxton11}
{Paxton}, B., {Bildsten}, L., {Dotter}, A., {et~al.} 2011, \apjs, 192, 3

\bibitem[{{Paxton} {et~al.}(2013){Paxton}, {Cantiello}, {Arras}, {Bildsten},
  {Brown}, {Dotter}, {Mankovich}, {Montgomery}, {Stello}, {Timmes}, \&
  {Townsend}}]{paxton13}
{Paxton}, B., {Cantiello}, M., {Arras}, P., {et~al.} 2013, \apjs, 208, 4

\bibitem[{{Paxton} {et~al.}(2015){Paxton}, {Marchant}, {Schwab}, {Bauer},
  {Bildsten}, {Cantiello}, {Dessart}, {Farmer}, {Hu}, {Langer}, {Townsend},
  {Townsley}, \& {Timmes}}]{paxton15}
{Paxton}, B., {Marchant}, P., {Schwab}, J., {et~al.} 2015, \apjs, 220, 15

\bibitem[{{Paxton} {et~al.}(2018){Paxton}, {Schwab}, {Bauer}, {Bildsten},
  {Blinnikov}, {Duffell}, {Farmer}, {Goldberg}, {Marchant}, {Sorokina},
  {Thoul}, {Townsend}, \& {Timmes}}]{paxton18}
{Paxton}, B., {Schwab}, J., {Bauer}, E.~B., {et~al.} 2018, \apjs, 234, 34

\bibitem[{{Paxton} {et~al.}(2019){Paxton}, {Smolec}, {Schwab}, {Gautschy},
  {Bildsten}, {Cantiello}, {Dotter}, {Farmer}, {Goldberg}, {Jermyn}, {Kanbur},
  {Marchant}, {Thoul}, {Townsend}, {Wolf}, {Zhang}, \& {Timmes}}]{paxton19}
{Paxton}, B., {Smolec}, R., {Schwab}, J., {et~al.} 2019, \apjs, 243, 10

\bibitem[{{Perryman} {et~al.}(1997){Perryman}, {Lindegren}, {Kovalevsky},
  {Hoeg}, {Bastian}, {Bernacca}, {Cr{\'e}z{\'e}}, {Donati}, {Grenon},
  {Grewing}, {van Leeuwen}, {van der Marel}, {Mignard}, {Murray}, {Le Poole},
  {Schrijver}, {Turon}, {Arenou}, {Froeschl{\'e}}, \& {Petersen}}]{perryman97}
{Perryman}, M.~A.~C., {Lindegren}, L., {Kovalevsky}, J., {et~al.} 1997, \aap,
  323, L49

\bibitem[{{Petrov} {et~al.}(2014){Petrov}, {Vink}, \&
  {Gr{\"a}fener}}]{petrov14}
{Petrov}, B., {Vink}, J.~S., \& {Gr{\"a}fener}, G. 2014, \aap, 565, A62

\bibitem[{{Prinja} {et~al.}(1990){Prinja}, {Barlow}, \& {Howarth}}]{prinja90}
{Prinja}, R.~K., {Barlow}, M.~J., \& {Howarth}, I.~D. 1990, \apj, 361, 607

\bibitem[{{Prinja} {et~al.}(2002){Prinja}, {Massa}, \& {Fullerton}}]{prinja02}
{Prinja}, R.~K., {Massa}, D., \& {Fullerton}, A.~W. 2002, \aap, 388, 587

\bibitem[{{Prinja} \& {Massa}(2010)}]{prinja10}
{Prinja}, R.~K. \& {Massa}, D.~L. 2010, \aap, 521, L55

\bibitem[{{Prinja} \& {Massa}(2013)}]{prinja2013}
{Prinja}, R.~K. \& {Massa}, D.~L. 2013, \aap, 559, A15

\bibitem[{{Przybilla} {et~al.}(2010){Przybilla}, {Firnstein}, {Nieva},
  {Meynet}, \& {Maeder}}]{przybilla10}
{Przybilla}, N., {Firnstein}, M., {Nieva}, M.~F., {Meynet}, G., \& {Maeder}, A.
  2010, \aap, 517, A38

\bibitem[{{Puls} {et~al.}(1996){Puls}, {Kudritzki}, {Herrero}, {Pauldrach},
  {Haser}, {Lennon}, {Gabler}, {Voels}, {Vilchez}, {Wachter}, \&
  {Feldmeier}}]{puls96}
{Puls}, J., {Kudritzki}, R.~P., {Herrero}, A., {et~al.} 1996, \aap, 305, 171

\bibitem[{{Puls} {et~al.}(2006){Puls}, {Markova}, {Scuderi}, {Stanghellini},
  {Taranova}, {Burnley}, \& {Howarth}}]{puls06}
{Puls}, J., {Markova}, N., {Scuderi}, S., {et~al.} 2006, \aap, 454, 625

\bibitem[{{Puls} {et~al.}(2020){Puls}, {Najarro}, {Sundqvist}, \&
  {Sen}}]{puls20}
{Puls}, J., {Najarro}, F., {Sundqvist}, J.~O., \& {Sen}, K. 2020, \aap, 642,
  A172

\bibitem[{{Puls} {et~al.}(2005){Puls}, {Urbaneja}, {Venero}, {Repolust},
  {Springmann}, {Jokuthy}, \& {Mokiem}}]{puls05}
{Puls}, J., {Urbaneja}, M.~A., {Venero}, R., {et~al.} 2005, \aap, 435, 669

\bibitem[{{Puls} {et~al.}(2008){Puls}, {Vink}, \& {Najarro}}]{puls08}
{Puls}, J., {Vink}, J.~S., \& {Najarro}, F. 2008, \aapr, 16, 209

\bibitem[{{Repolust} {et~al.}(2004){Repolust}, {Puls}, \&
  {Herrero}}]{repolust04}
{Repolust}, T., {Puls}, J., \& {Herrero}, A. 2004, \aap, 415, 349

\bibitem[{{Rivero Gonz{\'a}lez} {et~al.}(2011){Rivero Gonz{\'a}lez}, {Puls}, \&
  {Najarro}}]{rivero-gonzalez11}
{Rivero Gonz{\'a}lez}, J.~G., {Puls}, J., \& {Najarro}, F. 2011, \aap, 536, A58

\bibitem[{{Rubio-D{\'\i}ez} {et~al.}(2022){Rubio-D{\'\i}ez}, {Sundqvist},
  {Najarro}, {Traficante}, {Puls}, {Calzoletti}, \& {Figer}}]{rubio-diez22}
{Rubio-D{\'\i}ez}, M.~M., {Sundqvist}, J.~O., {Najarro}, F., {et~al.} 2022,
  \aap, 658, A61

\bibitem[{{Ryans} {et~al.}(2002){Ryans}, {Dufton}, {Rolleston}, {Lennon},
  {Keenan}, {Smoker}, \& {Lambert}}]{ryans02}
{Ryans}, R.~S.~I., {Dufton}, P.~L., {Rolleston}, W.~R.~J., {et~al.} 2002,
  \mnras, 336, 577

\bibitem[{{Sabhahit} {et~al.}(2022){Sabhahit}, {Vink}, {Higgins}, \&
  {Sander}}]{sabhahit22}
{Sabhahit}, G.~N., {Vink}, J.~S., {Higgins}, E.~R., \& {Sander}, A. A.~C. 2022,
  \mnras, 514, 3736

\bibitem[{{Santolaya-Rey} {et~al.}(1997){Santolaya-Rey}, {Puls}, \&
  {Herrero}}]{santolaya-rey97}
{Santolaya-Rey}, A.~E., {Puls}, J., \& {Herrero}, A. 1997, \aap, 323, 488

\bibitem[{{Sim{\'o}n-D{\'\i}az} \& {Herrero}(2014)}]{simon-diaz14a}
{Sim{\'o}n-D{\'\i}az}, S. \& {Herrero}, A. 2014, \aap, 562, A135

\bibitem[{{Skrutskie} {et~al.}(2006){Skrutskie}, {Cutri}, {Stiening},
  {Weinberg}, {Schneider}, {Carpenter}, {Beichman}, {Capps}, {Chester},
  {Elias}, {Huchra}, {Liebert}, {Lonsdale}, {Monet}, {Price}, {Seitzer},
  {Jarrett}, {Kirkpatrick}, {Gizis}, {Howard}, {Evans}, {Fowler}, {Fullmer},
  {Hurt}, {Light}, {Kopan}, {Marsh}, {McCallon}, {Tam}, {Van Dyk}, \&
  {Wheelock}}]{skrutskie06}
{Skrutskie}, M.~F., {Cutri}, R.~M., {Stiening}, R., {et~al.} 2006, \aj, 131,
  1163

\bibitem[{{Smith}(2014)}]{smith14}
{Smith}, N. 2014, \araa, 52, 487

\bibitem[{{Sobolev}(1960)}]{sobolev60}
{Sobolev}, V.~V. 1960, {Moving Envelopes of Stars}

\bibitem[{{Sundqvist} \& {Owocki}(2013)}]{sundqvist13}
{Sundqvist}, J.~O. \& {Owocki}, S.~P. 2013, \mnras, 428, 1837

\bibitem[{{Sundqvist} \& {Puls}(2018)}]{sundqvist18}
{Sundqvist}, J.~O. \& {Puls}, J. 2018, \aap, 619, A59

\bibitem[{{Sundqvist} {et~al.}(2014){Sundqvist}, {Puls}, \&
  {Owocki}}]{sundqvist14}
{Sundqvist}, J.~O., {Puls}, J., \& {Owocki}, S.~P. 2014, \aap, 568, A59

\bibitem[{Townsend(2020)}]{sdk20}
Townsend, R. 2020, MESA SDK for Linux

\bibitem[{{Vink}(2022)}]{vink22}
{Vink}, J.~S. 2022, \araa, 60, 203

\bibitem[{{Vink} {et~al.}(2010){Vink}, {Brott}, {Gr{\"a}fener}, {Langer}, {de
  Koter}, \& {Lennon}}]{vink10}
{Vink}, J.~S., {Brott}, I., {Gr{\"a}fener}, G., {et~al.} 2010, \aap, 512, L7

\bibitem[{{Vink} {et~al.}(1999){Vink}, {de Koter}, \& {Lamers}}]{vink99}
{Vink}, J.~S., {de Koter}, A., \& {Lamers}, H.~J.~G.~L.~M. 1999, \aap, 350, 181

\bibitem[{{Vink} {et~al.}(2001){Vink}, {de Koter}, \& {Lamers}}]{vink01}
{Vink}, J.~S., {de Koter}, A., \& {Lamers}, H.~J.~G.~L.~M. 2001, \aap, 369, 574

\bibitem[{{{\v{S}}urlan} {et~al.}(2012){{\v{S}}urlan}, {Hamann}, {Kub{\'a}t},
  {Oskinova}, \& {Feldmeier}}]{surlan12}
{{\v{S}}urlan}, B., {Hamann}, W.~R., {Kub{\'a}t}, J., {Oskinova}, L.~M., \&
  {Feldmeier}, A. 2012, \aap, 541, A37

\end{thebibliography}


\begin{appendix}
%
\section{Stellar parameter determination in a nutshell}
\label{apen.stellar_param}

The line-broadening analysis of the stars in the sample has been done using the {\tt IACOB-BROAD} tool \citep[see][]{simon-diaz14a}, which allowed us to derive estimates of \vsini.
The quantitative spectroscopic analysis of the stars in this work was performed using a grid of unclumped model atmospheres computed with the NLTE model atmosphere and line synthesis \textsc{} Fast Analysis of STellar
atmospheres with WINDs code \citep[FASTWIND, v10.4.7,][]{santolaya-rey97, puls05, rivero-gonzalez11, puls20}. These models were used with supervised learning techniques to create a statistical emulator for \textsc{FASTWIND} synthetic spectra that, in combination with a Markov chain Monte Carlo (MCMC) method, allowed us to obtain estimates of \Teff, \logg, and $Q$ \citep[see][Sect.~3 for further details]{deBurgos24a}. 
We note that for a small group of stars, their H$\alpha$ profiles exhibit a small double-subpeak structure reaching the continuum level. For precaution, these stars are identified with a different symbol in all figures.

We adopted geometric distances from \citet{bailer-jones21} using \textit{Gaia} EDR3 data \citep{gaiacollaboration21}. The corresponding error-over-parallax for 95\% of our sample takes values \ls0.15, being 0.15\,--\,0.20 for the remaining 5\%. For some very bright stars ($G_{mag}\ls4$) initially considered in the sample, we also searched for alternative parallax values in \textit{Hipparcos} \citep{perryman97}. However, all of them had associated error-over-parallax values greater than 0.20 and we did not include them.

We used adopted distances combined with optical photometry ($B$, $V$) from \citet{mermilliod06} and infrared photometry ($J$, $H$, $Ks$) from \citet[][\textit{2MASS}]{skrutskie06} to derive the extinction ($A_{\rm V}$) to each star. In particular, our methodology is based on an MCMC to solve simultaneously for the color excess $E(B-V)$ and $R_\mathrm{v}$, by comparing the observed spectral energy distribution (SED) of the star with the one predicted by \textsc{FASTWIND} for the corresponding stellar parameters. 

Combining the estimated $A_{\rm V}$ with the collected apparent $V$ magnitudes and distances, we were able to first compute the absolute $V$ magnitudes and then, following \citet{herrero92}, the stellar radii. In a second step, and also taking into account the effective temperatures, surface gravities, projected rotational velocities, and wind-strength parameters resulting from the quantitative spectroscopic analysis, we computed the stellar luminosities, the surface gravities corrected from centrifugal forces \citep[following][]{repolust04} and, eventually, for those stars with available information about the wind terminal velocities, the associated mass-loss rates.

%
\section{Uncertainties in the parameters}
\label{apen.uncertainties}
The uncertainties of the spectroscopic parameters \Teff, \logg, and \logQ, as well as those of $A_{\rm V}$, were taken as 33\% on both sides of the maximum of the corresponding probability distribution function \citep[see][for more details]{deBurgos24a}. Regarding distances, we averaged the 16th and 84th percentiles provided in \citet{bailer-jones21}. Following \citet{prinja90}, we adopted a global uncertainty of 100\kms for \vinf. For the remaining parameters ($R$, $L$, and $\dot{M}$), we adopted the standard deviation resulting from a Monte Carlo approach that consisted of generating 1000 Gaussian random values within the uncertainties of the parameters required in each calculation.

%
\section{Evolutionary models}
\label{apen.evo}

For our numerical calculations of 1D stellar evolution models, we used release r22.11.01 of the Modules for Experiments in Stellar Astrophysics, \textsc{mesa}, software instrument \citep{paxton11, paxton13, paxton15, paxton18, paxton19}. \textsc{mesa} is an open-source, robust tool, well-tested in modeling mass-loss rates from massive stars \citep[e.g.,][]{keszthelyi17, keszthelyi22, higgins21, sabhahit22}. 

We used the \textsc{mesa} Software Development Kit (\textsc{SDK}) version 22.05.01 \citep[][]{sdk20}, and we performed our calculations on the calculation server of the Japanese supercomputer XC50 \footnote{\url{https://https://cfca.nao.ac.jp/}}. We calculated models with initial masses in the range of 20~to~60\MSol, with 5\MSol increments covering the evolutionary phase from the ZAMS until 10\,kK. We stopped the calculations here because it is already 5\,kK below the lower boundary covered by the observational sample. At 10\,kK, some models are still on the main sequence, whereas others have depleted hydrogen in their cores and are in their post-main sequence evolution. 

We assumed the Galactic chemical composition with initially $Z = 0.014$ and the \citet{asplund09} metal fractions modified for certain elements by \citet{przybilla10} and \citet{nieva12}. The isotopic ratios were taken from \citet{lodders03}. The initial hydrogen and helium mass fractions were adopted as $X=0.720$ and $Y=0.266$, respectively.

We assumed a mixing length parameter of $\alpha_{\rm MLT} = 1.8$ and the ``ML1" option in \textsc{mesa}, adopting the \citet{bohm58} description of convective mixing. We used exponential overshooting at convective boundaries with $f=0.025$ and $f0=~0.005$. Very approximately, this corresponds to an extension by 20\% of the local pressure scale height.

The initial rotation was set to a value of $\Omega/\Omega_{\rm crit} = 0.2$ and rotational mixing and angular momentum transport are assumed as diffusive processes. Similarly to \citet{vink10}, a strong core envelope coupling is adopted in the models by means of the viscosity of the Spruit-Tayler dynamo.

The mass-loss rate by stellar winds was calculated using three different theoretical predictions. First, we calculated a branch of models adopting the \citet{vink01} rates (their Eqs. 24-25). Stellar evolution codes can have some differences in their implementations, even when formally the same prescription is used. This is because the bistability region requires interpolation (from \citealt{vink01}, Eq. 24 is used for \Teff\,>\,27.5\,kK, whereas Eq. 25 is used for \Teff\,<\,22.5\,kK), which is not provided in the prescription (see the discussion by \citealt{keszthelyi17}). Here, we chose to use the routine directly implemented in \textsc{mesa}, which sets the bistability temperature based on metallicity (Eqs. 14-15 of \citealt{vink01}), and uses a narrow 1\,kK interpolation range. This is similar to the implementation of \citet{ekstrom12} but differs from the more gradual change implemented by \citet{brott11}. Besides metallicity, the location of the bistability also depends on the Eddington factor ($\Gamma_{e}$; Eq. 23 of \citealt{vink01}). However, the effect of $\Gamma_{e}$ only produces a small shift of less than 2\,kK in the predicted \Teff\ values where the bistability jump would occur.
Second, we implemented the mass-loss prescription obtained by \citet{krticka21} (their Eq. 3). This formula also accounts for an increase in the mass-loss rates over the bistability region, though at a much lower effective temperature.
Third, we calculated a branch of models where the hot star mass-loss rates utilize the prescription of \citet{bjorklund23} (their Eq. 7). 

For the latter two branches, we modified the \texttt{run\_star\_extras} file to calculate and apply mass-loss rates that are not directly included in \textsc{mesa}. Since the theoretical mass-loss rates do not account for the effects of rotation, we scaled the nominal mass-loss rates by rotational enhancement, following the formula of \citet{maeder00}. Since rotation is slow in the models, this results in a negligibly small enhancement ($1\%$). We should note, however, that the base solar metallicity is adopted differently in these wind prescriptions. \citet{vink01} use $Z_\odot$\,=0.019, whereas \citet{krticka21} and \citet{bjorklund23} adopt $Z_\odot$\,=0.0134.
Since a metallicity scaling is included in all prescriptions, this can also lead to some differences between calculations formally at "solar" metallicity.

The calculation of wind terminal velocities \vinf\ is independent of the mass-loss rate calculations in the evolutionary models. We simply set empirical constraints and calculate \vinf\ from the escape velocity, $v_{\rm esc} = \sqrt{(1-\Gamma_{e}) 2 G M / R} $. Namely, we adopted $v_\infty / v_{\rm esc}$\,=3.0 and 1.3 for \Teff\ larger than and smaller than 22\,kK, respectively. The chosen interpolation width is 3\,kK.

%
\section{Summary values of mass-loss rates and wind-strength parameter}
\label{apen.meanval}

Table~\ref{tab.avgQMdot} summarizes the average value and standard deviation of $Q$ and $\dot{M}$ for stars grouped in bins of \logL\ and \Teff. We decided to exclude those stars for which the estimates of these parameters are upper or lower limits (see Fig.~\ref{fig.fig1}), thus not representative of the real value. For comparison, Table~\ref{tab.avgQMdot} also includes the results using the full sample (values in parentheses).

For stars in the highest luminosity bin, the average values without parentheses clearly show a decrease in $Q$ and $\dot{M}$ with \Teff. In fact, this decrease is three and two times greater than the average scatter, respectively. For stars in the middle luminosity bin, both quantities evolve practically flat with \Teff, at an average value of $-$13.4\,dex for $Q$, and $-$6.83\,dex for $\dot{M}$. For $Q$, we note that the average values are slightly lower above 25\,kK than below, but the scatter values are consistent with a flat trend. Lastly, the values and scatter of stars in the lowest luminosity bin also reflect a flat trend in both quantities. We note that for stars above 30\,kK, most of the stars represent upper- and lower-limit cases of $Q$. Therefore, we observe the largest difference with the values in parentheses.

\begin{table*}[!t]
 \centering
 \caption{Average and standard deviation of the wind-strength values and mass-loss rates in the sample grouped in bins of \logL\ and \Teff.}
 \label{tab.avgQMdot}
  \begin{tabular}{cc|cccc}
  \hline
  \hline
     \noalign{\smallskip}
    \multicolumn{2}{c|}{\multirow{2}{*}{\logQ}} & \multicolumn{4}{c}{\Teff\,[kK]} \\
    &  & 35\,--\,30 & 30\,--\,25 & 25\,--\,20 & 20\,--\,15 \\
     \hline
  \multirow{6}{*}{\rotatebox{90}{\logL}} & \multirow{2}{*}{5.3\,--\,5.8} 
     & $-$12.69$\pm$0.22 & $-$13.05$\pm$0.21 & $-$13.13$\pm$0.15 & $-$13.23$\pm$0.14 \\
  &  &($-$12.65$\pm$0.18)&($-$13.05$\pm$0.21)&($-$13.13$\pm$0.15)&($-$13.23$\pm$0.14)\\
  & \multirow{2}{*}{5.0\,--\,5.3} 
     & $-$13.41$\pm$0.34 & $-$13.51$\pm$0.19 & $-$13.31$\pm$0.08 & $-$13.33$\pm$0.07 \\
  &  &($-$13.33$\pm$0.37)&($-$13.54$\pm$0.21)&($-$13.48$\pm$0.31)&($-$13.33$\pm$0.07)\\
  & \multirow{2}{*}{4.7\,--\,5.0}
     & $-$13.27$\pm$0.27 & $-$13.45$\pm$0.28 & $-$13.46$\pm$0.24 & ... \\
  &  &($-$13.74$\pm$0.34)&($-$13.63$\pm$0.33)&($-$13.65$\pm$0.31)& ... \\
    \hline
    \multicolumn{2}{c|}{\multirow{2}{*}{\logMdot}} & \multicolumn{4}{c}{\Teff\,[kK]} \\
    &  & 35\,--\,30 & 30\,--\,25 & 25\,--\,20 & 20\,--\,15 \\
     \hline
  \multirow{6}{*}{\rotatebox{90}{\logL}} & \multirow{2}{*}{5.3\,--\,5.8}
    & $-$5.96$\pm$0.39 & $-$6.17$\pm$0.22 & $-$6.40$\pm$0.24 & $-$6.51$\pm$0.12 \\
  & &($-$5.86$\pm$0.32)&($-$6.17$\pm$0.22)&($-$6.40$\pm$0.24)&($-$6.51$\pm$0.12)\\
  & \multirow{2}{*}{5.0\,--\,5.3}
    & $-$6.87$\pm$0.29 & $-$6.91$\pm$0.18 & $-$6.71$\pm$0.07 & $-$6.81$\pm$0.14 \\
  & &($-$6.81$\pm$0.30)&($-$6.93$\pm$0.21)&($-$6.88$\pm$0.31)&($-$6.81$\pm$0.14)\\
  & \multirow{2}{*}{4.7\,--\,5.0}
    & $-$7.17$\pm$0.10 & $-$7.03$\pm$0.26 & $-$7.09$\pm$0.43 & ... \\
  & &($-$7.50$\pm$0.22)&($-$7.20$\pm$0.30)&($-$7.22$\pm$0.38)& ... \\
    \hline
 \end{tabular}
 \tablefoot{The values in parenthesis take into account all the 116 stars in the sample. The values without parentheses exclude stars for which the estimates of $Q$ and $\dot{M}$ represent upper or lower limits.}
\end{table*}

%
\section{Wind terminal velocities}
\label{apen.vinf}

\begin{figure}[!t]
    \centering
    \includegraphics[width=0.48\textwidth]{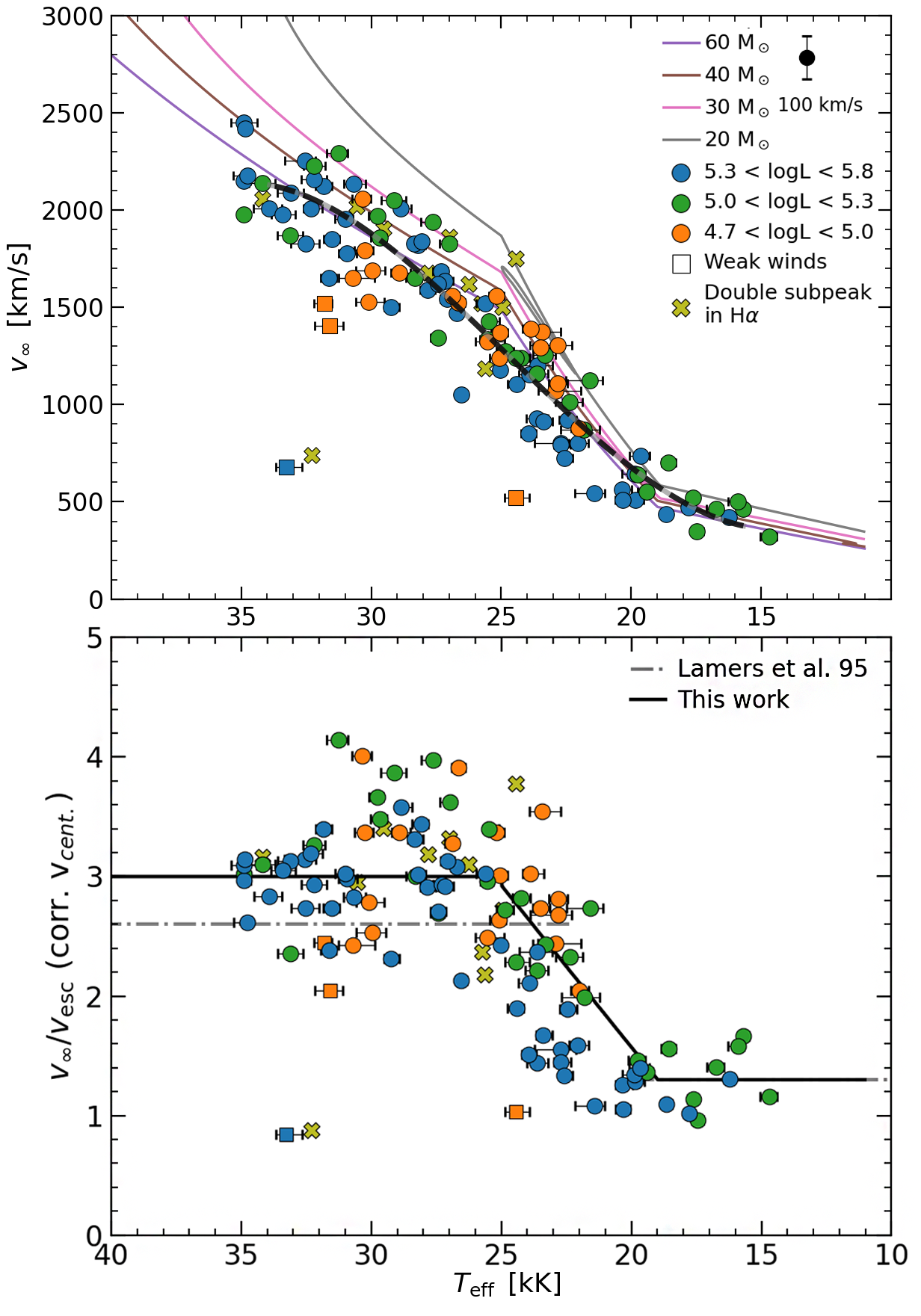}
    \caption{Stars in the sample colored by different derived quantities against their effective temperature.  
    Top panel: Adopted values of wind terminal velocities from the literature, whereas the bottom panel shows \vinf/\vesc, where estimates of \vesc\ have been corrected from centrifugal forces. The tracks and symbols are the same as in Fig.\ref{fig.fig1}. The average uncertainty in \vinf\ is indicated as an error-bar within the legend. The black solid line corresponds to a third order regression of those stars shown with filled circles.
    In the bottom panel, the solid line indicates our adopted prescription for \vinf/\vesc, whereas the dashed-dotted line indicates the prescription by \citet{lamers95}.}
\label{fig.fig_vinf_vesc}
\end{figure}

The top panel of Fig.~\ref{fig.fig_vinf_vesc} shows our adopted wind terminal velocities from the literature against our derived effective temperatures for the same stars as in Fig.\ref{fig.fig1}.
For completeness, we decided to performed a third-order regression fit to the data, which is also shown in the panel, and follows the equation:
\begin{align*}
v_{\infty}~[km/s] = -0.416T_{\rm eff}^3 +30.9 T_{\rm eff}^2 -633 T_{\rm eff} +4310~[kK].
\end{align*}
We notice that compared to the models, the drop in \vinf\ below \Teff\,$\approx$25\,kK becomes very smooth in the observational data, as also shown by the regression fit.

The same panel also shows a group of stars that have lower terminal velocities compared to the other stars of the same luminosity range. Those located at \vinf\,\ls1500\,\kms and \Teff\,>\,30\,kK correspond to HD\,76341 (O9.2\,IV), HD\,163892 (O9.5\,IV(n)), HD\,191423 (ON9\,II-IIInn), and HD\,306097 (O9\,V) from larger to lower \vinf. They belong to a group of stars frequently called ``the weak-wind O stars" \citep[see][]{martins05b, marcolino09, deAlmeida19}, and are characterized for being O9 III\,--\,V type stars with lower mass-loss rates compared to most O-type stars. An additional star at \vinf\,$\approx$500\,\kms and 25\,kK is HD\,86606 (B1\,Ib). Its spectral type and unusually low \vinf\ makes it a strong candidate to be a magnetic star and certainly demands further study.

%
\section{\textsc{FASTWIND} simulations with clumping}
\label{apen.macroclump}

\begin{figure}[!t]
    \centering
    \includegraphics[width=0.48\textwidth]{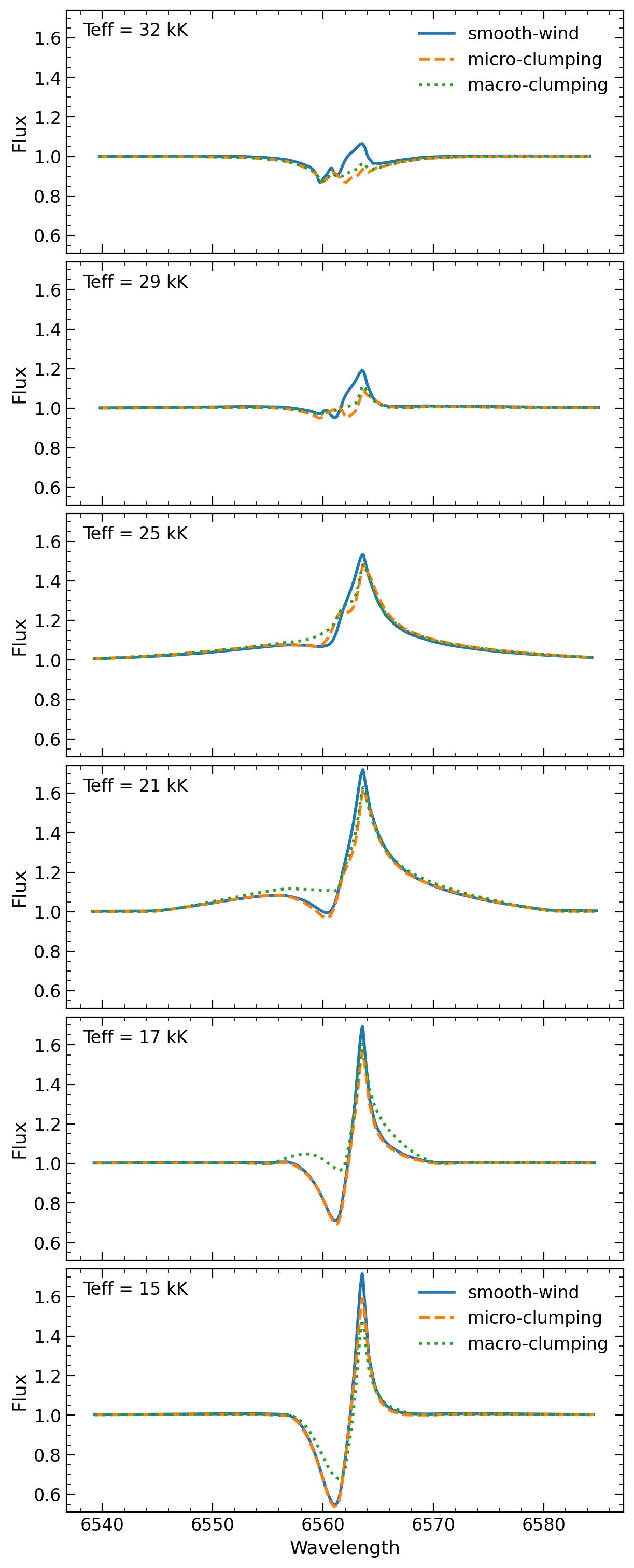}
    \caption{Effect of micro- and macro-clumping in the H$\alpha$ line. Each sub-panel corresponds to a different temperature and includes three models for different wind treatments: Smooth wind or unclumped (solid blue line), wind with micro-clumping (dashed orange line), and wind with macro-clumping for $f_\mathrm{ic}$\,=0.2 (dotted green line).}
    \label{fig.test_macroclump}
\end{figure}

We used \textsc{FASTWIND} (v10.6.5.1) to simulate several sequences of models for constant luminosity \logL\,=5.4\,dex, representative of the bulk of the sample of stars studied in this work. The sequences span the 40\,--\,15\,kK range in effective temperature (with a step of 1\,kK), for eight different values of the wind-strength parameter defined by $\log{Q_\mathrm{s}}$ (where the index {\em s} indicates the value of a smooth wind), encompassing the range of values derived in our study. The surface gravity for each temperature was selected to be representative of the corresponding value for the luminosity; the radii follow from the luminosity and the temperatures, with the wind terminal velocity related to the escape velocity by the relationship provided in \citet{kudritzki00}. For all models, as a compromise, we adopted a fixed value for the exponent of the wind velocity law, $\beta$\,=1.5. Solar abundances were adopted for all species, including helium.  
For a given $\log{Q_\mathrm{s}}$ sequence, three sets of models were calculated: (1) smooth wind; (2) optically thin clumps considering a clumping factor $f_\mathrm{cl}$\,=10 (i.e., volume filling factor of 0.1), following a linear increase law with clumping starting at 0.03$v_\infty$ and reaching its maximum value at 0.1$v_\infty$; and (3) macro-clumping in the wind, considering the following values \citep[see][for the detailed discussion of their meaning]{sundqvist18}: clumping factor $f_\mathrm{cl}$\,=10, interclump density contrast $f_\mathrm{ic}$\,=0.01 and 0.2 \citep[guided by the results obtained by][]{hawcroft21, brands22}, velocity filling factor $f_\mathrm{vel}$\,=0.5. and porosity length at the wind terminal velocity $h$\,=1 (in units of stellar radius). For models considering clumping, the mass-loss rates were scaled with the clumping factor according to $\dot{M}_\mathrm{cl} = \dot{M}_\mathrm{s}\,f_\mathrm{cl}^{-1/2}$, regardless of the nature of the clumping. In all cases, we account for the effect of non-coherent electron scattering in the calculation of the emergent profiles.   

Fig.~\ref{fig.test_macroclump} illustrates the outcome of these simulations for a particular value of $\log{Q_\mathrm{s}}=-$12.7\,dex for a selected number of models. The conclusions regarding the behavior of H$\alpha$ are very similar for all other $\log{Q_\mathrm{s}}$ sequences.
As can be readily seen, the simulations show that the mass-loss rates that would be derived by applying models that do not consider any clumping would still represent a maximum mass-loss rate limit. That is, regardless of the nature of the clumping, the mass-loss rates derived using clumped models would always be smaller than those derived from smooth-wind models.
Therefore, this allows us to conclude that (at least for winds of early to mid B-type supergiants of luminosity class Ia or lower) there is no indication whatsoever of an increased mass-loss rate that would correspond to the so-called bistability jump predicted in the work of \citet{vink01}.

\end{appendix}

\end{document}